\documentclass{article}
\usepackage{spconf,amsmath,graphicx,booktabs,xcolor,hyperref}
\usepackage{siunitx}

\usepackage{amsmath,amsfonts,bm}

\def\eqref#1{equation~\ref{#1}}

\def\1{\bm{1}}

\def\va{{\bm{a}}}
\def\vb{{\bm{b}}}
\def\vc{{\bm{c}}}

\def\vm{{\bm{m}}}

\def\mS{{\bm{S}}}

\DeclareMathAlphabet{\mathsfit}{\encodingdefault}{\sfdefault}{m}{sl}
\SetMathAlphabet{\mathsfit}{bold}{\encodingdefault}{\sfdefault}{bx}{n}

\title{Semi-supervised Monaural Singing Voice Separation with\\a Masking Network Trained on Synthetic Mixtures}
\name{Michael Michelashvili$^1$, Sagie Benaim$^1$,  Lior Wolf$^{1,2}$}
\address{$^{1}$Tel Aviv University~~~~~~~~~~ $^{2}$Facebook AI Research}
\allowdisplaybreaks
\begin{document}
%
\maketitle
%

\begin{abstract}
We study the problem of semi-supervised singing voice separation, in which the training data contains a set of samples of mixed music (singing and instrumental) and an unmatched set of instrumental music. 
Our solution employs a single mapping function $g$, which, applied to a mixed sample, recovers the underlying instrumental music, and,  applied to an instrumental sample, returns the same sample. The network $g$ is trained using purely instrumental samples, as well as on synthetic mixed samples that are created by mixing reconstructed singing voices with random instrumental samples. Our results indicate that we are on a par with or better than fully supervised methods, which are also provided with training samples of unmixed singing voices, and are better than other recent semi-supervised methods.
\end{abstract}

\section{Introduction}
\label{sec:intro}

The problem of separating a given mixed signal into its components without direct supervision is ubiquitous. For example, in single cell gene expression conducted in cancer research, one obtains a gene expression that contains both the cancer cell of interest and the expression of immune cells that attach to it. In what is known in biology as gene expression deconvolution~\cite{zhao2010gene}, one would like to obtain the expression of the cancer cell itself, while only having access to a dataset of such mixed readings and another dataset containing gene expression profiles of immune cells. 

In the task of singing voice separation, which is the focus of this work, examples of mixed music, which contains both singing and instrumental music, are abundant. It is also relatively easy to label parts of the song where no singing is present. However, it is much  harder to separate out pure voice samples. Without such samples, one cannot use the supervised methods that were suggested for this separation task. 

In this work, we propose a novel method for performing the separation. 
The method is based on applying a learned function twice: once on the mixtures, in order to recover estimated singing voice samples, and once on synthetic mixes, in which the reconstructed singing samples are crossed with real instrumental samples from the training set. The advantage of these crosses over the original mixed samples is that the underlying components of these mixed samples are known, and, therefore,  added losses can be applied, when training the separating function on them. 

\section{Related Work}
\label{sec:prev}
%
Single-channel source separation is a long-standing task which has been researched extensively. Classical works on blind  source separation include Single-Channel ICA \cite{davies2007source} and, specifically in singing voice separation, RPCA~\cite{jeo2}. These methods utilize hand-crafted priors on the sources, such as a low rank assumption on the instrumental music. 

The problem of singing voice separation is often studied in the supervised case, where the mixed samples are provided with the target source. Often, a simple masking model in the spectral domain is assumed and the desired source $\vb$ is given by a point-wise  multiplication of the mixed signal $\va$ and some mask $\vm$, i.e., $\vb = \va \odot \vm$, where $\odot$ is the Hadamard product. In our work, we use a network $g$ such that $\vb \approx \va-g(\va)$, where the architecture of $g$ includes the masking, i.e., $g(\va) = \va  \odot m(\va)$, for some subnetwork $m$ with outputs in $[0,1]$.

The GRA3 method~\cite{gra3}, similarly to ours, estimates the mask $\vm$ directly from the mixed sample $\va$. This is done using an ensemble of four deep neural networks, trained with different losses. The architecture we use is of the type commonly used for the autoencoding of images. Similar architectures are used in other work to directly estimate all the sources from the mixtures, e.g.,~\cite{7178348,chandna2017monoaural}. 

The GRU-RIS-L method of Mimilakis et al.~\cite{mimilakis2018monaural}, employs RNNs of stochastic depth in order to recover the time-frequency mask. The usage of RNNs allows for efficient modeling of longer time dependencies of the input data.  This is extended in~\cite{madtwin} (MaDTwinNet) by introducing a technique called Twin Net, which regularizes the RNNs.
Our analysis of long sequences is segment by segment and does not exploit long range dependencies. 

Adversarial training using GANs~\cite{goodfellow2014generative} is a powerful method for unconditional image generation. GANs are composed of two parts: (i) a generator $g$ that synthesizes realistic images, and (ii) a discriminator $d$ that distinguishes real from fake images. The objective of the generator is to create images that are realistic enough to fool the discriminator. The objective of the discriminator is to detect the fake images. The method was later extended to perform unsupervised image-to-image mapping~\cite{CycleGAN2017}. In this setting, the generator is conditioned on an input image from the source domains and generates a ``fake'' sample in the target domain. As in the unconditional setting, the discriminator attempts to differentiate between real and generated images. Adversarial training was used for supervised source separation, where the distribution of each of the mixture components is known and modeled by a GAN, by  Stoller et al.~\cite{stoller2017adversarial} and Subkhan et al.~\cite{subakan2017generative}. The adversarial training was motivated as being better able to deal with correlated sources. Semi-supervised approach using adverserial training was used by Higuchi et al \cite{@Higuchi2017} for the task of speech enhancement.

In the setting of Semi-supervised audio source separation, in which we work, the task is to separate mixtures of two sources given mixed samples as well as samples from only one of the sources. Previous solutions were typically based on NMF~\cite{barker2014semi} or the related PLCA~\cite{smaragdis2007supervised}. 

The most similar method to ours is NES~\cite{yedidiclr2019}, which separates mixed samples into a sum of two samples: one from an observed domain and one from an unobserved domain. The method consists of an iterative process: (i) estimation of samples from the unobserved distribution; (ii) synthesis of mixed signals by combining training samples from the observed domain  and the estimated samples from the unobserved one; (iii) training of a mapping from the mixed domain to the observed domain. 
It was demonstrated in~\cite{yedidiclr2019} that due to its iterative nature, NES is sensitive to the initialization method. Our method, in contrast, performs a non-iterative end to end training that includes the synthetic mixtures as part of the network. This also allows us to apply additional losses, such as GAN based losses and the constraint that learned function $g$ is idempotent ($g\circ g = g$)~\cite{galanti2017theory}. As can be seen in Sec.~\ref{sec:exp}, our results are significantly stronger than those obtained by~\cite{yedidiclr2019}.


\section{Method}
\label{sec:method}

\begin{figure}[t]
\centering
{\includegraphics[width=0.445\textwidth]{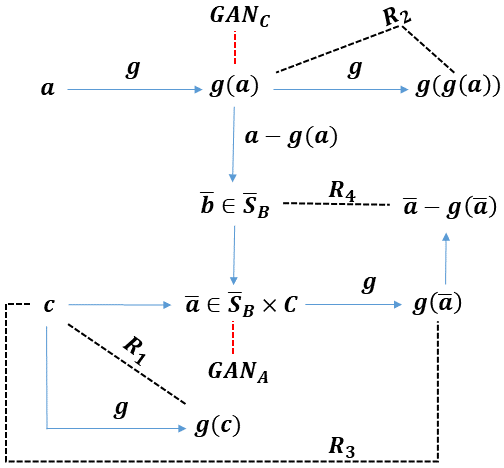}}
\caption{The transformations and constraints of our method. Blue arrows stand for functions. Dashed lines represent losses, which are of two types: reconstruction losses (black) and GAN loss terms (red).} 
	\label{fig:arch}
\end{figure}

In the problem of semi-supervised separation, the learning algorithm is provided with unlabeled datasets from two domains, a domain of mixtures $A$ and a domain of observed components $C$. There also exists a target domain $B$, from which no samples are presented. The goal is to learn a function $g:A\rightarrow C$, which maps a sample in domain $\va\in A$ to a component $\vc$ in domain $C$ such that there exits a component $\vb\in B$ for which the following equality holds $\va=\vb+\vc$.

During training, we obtain two sets of unmatched samples: the set $\mS_A$ of mixed samples in domain $A$, and the set $\mS_C$ of samples in the observed domain $C$.

Due to the lack of training samples in $B$, we rely on the generation of a synthetic training set of samples in domain $B$: 

The network mixes the samples in $\bar \mS_B$ with random samples in $C$, in order to create the following set of synthetic crosses:

For each sample $\bar \va \in \bar \mS_B \times C$, we memorize the underlying samples $\bar \vb,\vc$ that were used to create it, and mark these samples as $b(\bar \va)$ and $c(\bar \va)$, respectively.

In addition to  $g$, we train two discriminator networks $d_C$ and $d_A$, which provide adversarial signals that enforce the distribution of the recovered samples from domain $C$ to match the distribution of the training set $\mS_C$ and the mixed synthetic samples to match the distribution of domain $\mS_A$. Specifically, $d_C$ is applied to samples of the form $g(\va)$, where $\va \in \mS_A$; $d_A$ is applied to samples of the form $\bar \va \in \bar \mS_B \times C$.

The following losses are used to train the network $g$:
\begin{align}
\mathcal{L}_{R_1} =&  \sum_{\vc \in \mS_C} \|g(\vc)- \vc\|_1  \\ 
\mathcal{L}_{R_2} =&  \sum_{\va \in \mS_A} \|g(g(\va))- g(\va)\|_1  \\ 
\mathcal{L}_{R_3} =&  \sum_{\bar \va \in \bar \mS_B \times C} \|g(\bar \va)- c(\bar \va)\|_1\\   
\mathcal{L}_{R_4} =&  \sum_{\bar \va \in \bar \mS_B \times C} \|(\bar \va - g(\bar \va))- b(\bar \va)\|_1 \\
\mathcal{L}_{\text{GAN}_C} =& \sum_{\va \in \mS_A} -\ell(d_C(g(\va)),0)\\ 
\mathcal{L}_{\text{GAN}_A} =& \sum_{\bar \va \in \bar \mS_B \times C} -\ell(d_A(\bar \va),0),
\end{align}
where $\ell$ is the Least Squares loss, following~\cite{mao2017least}. That is, $\ell(x, y) = (x-y)^2$. Note that $g$ appears in $\mathcal{L}_{\text{GAN}_A}$ and appears more than once in $\mathcal{L}_{R_3}, \mathcal{L}_{R_4}$, since it takes part in the formation of the set $\bar \mS_B \times C$. 

The first loss requires that $g$, applied to samples in $C$, is the identity operator. The second loss enforces idempotence on $g$ (since $g$ maps to domain $C$, applying it again should be the same as applying identity), and the next two losses enforce the separation of the synthetic cross samples to result in the known components. The last two losses are GAN based losses in the domains C and A. 
The full objective for $g$ is defined as:
\begin{align*}
\mathcal{L}_g = \mathcal{L}_{R_1} + \mathcal{L}_{R_1} 
+ \mathcal{L}_{R_3} + \mathcal{L}_{R_4} 
+ 0.5(\mathcal{L}_{GAN_C} + \mathcal{L}_{GAN_A})
\end{align*}

The discriminators of the GAN losses, $d_C$ and $d_A$, are trained with the following losses, respectively:
\begin{align}
\mathcal{L}_{{d}_C} =& \sum_{\va \in \mS_A} \ell(d_C(g(\va)),0) + \sum_{\vc \in \mS_C} \ell(d_C(\vc),1)\\
\mathcal{L}_{{d}_A} =& \sum_{\bar \va \in \bar \mS_B \times C} \ell(d_A(\bar \va),0) + \sum_{\va \in \mS_A} \ell(d_A(\va),1) 
\end{align}

\section{Implementation Details}

An Adam optimizer is used with $\beta_1 = 0.5$, $\beta_2 = 0.999$ and a batch size of one. The learning rate is initially set to $0.0001$ and is halved after $100,000$ iterations. 

\subsection{Network architecture}
\label{sec:arch}

The underlying network architecture adapts that used in \cite{munit}. Let $C7S1_k$ denote a $7 \times 7$ 1-stride convolution with $k$ filters. Similarly, let $C4S2_k$ denote a $4 \times 4$ 2-stride convolution with $k$ filters. Let $R_k$ denote a residual block with two $3 \times 3$ convolutional blocks and $k$ filters and let $u_k$ denote a $2 \times$nearest-neighbor upsampling layer, followed by a $5 \times 5$ convolutional block with $k$ filters and 1 stride. 

Recall that $g(\va) = \va  \odot m(\va)$. $m$ is built as an auto-encoder. The encoder consists of two downsampling convolutional layers, $C7S1_{64}$ and $C4S2_{128}$. This is followed by four residual blocks of type $R_{256}$. Each convolutional layer of the encoder is followed by an Instance Normalization layer and a ReLU activation. The decoder consists of four residual blocks of type $R_{256}$. This is followed by two upsampling blocks, $u_{128}$ and $u_{256}$, and a convolutional layer $C7S1_3$. Each convolutional layer of the decoder is followed by a Adaptive Instance Normalization~\cite{adaptiveinstance} layer and a ReLU activation. To obtain mask values between $0$ and $1$, the ReLU of the last layer is replaced by a sigmoid activation function.

A multi-scale discriminator is used for $d_C$ and $d_A$, as in \cite{pix2pixhd}, to produce both accurate low-level details, as well as capture global structure. Each discriminator consists of  the following sequence of layers: $C4S2_{64}$, $C4S2_{128}$, $C4S2_{256}$ and $C4S2_{512}$. Each convolutional layer is followed by a leaky ReLU with slope parameter of $0.2$.

\subsection{Audio processing}
\label{sec:prep}
To convert an audio file to an input to network $g$, we perform the following pre-processing: The audio file is re-sampled to $20480$ Hz. It is then split into clips of duration of $0.825$ seconds. We then compute the Short Time Fourier Transform (STFT) with window size of $40$ms, hop size of $64$ and FFT size of $512$, resulting in an STFT of size $257 \times 256$. Lastly, we take the absolute values and apply a power-law compression with $p=0.3$, i.e. we obtain $|A|^{0.3}$, where $|A|$ is the magnitude of the STFT. The highest frequency bin is trimmed, resulting in an input audio representation of size $256 \times 256$.  

To convert the method's output $\bar \vb = \va-g(\va)$ back to audio, 
we apply ISTFT on the multiplication of the magnitude spectrogram of $\bar \vb$ with the phase of the original mixture, and add back the top-frequency by padding with zeros. To process an entire audio file, we simply process each non overlapping segment individually, and concatenate the results.

\section{Evaluation}
\label{sec:exp}

We perform a comparison to other semi-supervised methods, using the evaluation protocol used by~\cite{yedidiclr2019}. In addition, we compare our semi-supervised method to the state of the art supervised methods, following the protocol used in~\cite{madtwin}. Finally, ablation experiments are run to study the relative importance of the various losses.

\subsection{Comparison to semi-supervised methods}

For semi-supervised methods, our evaluation protocol follows closely the one of~\cite{yedidiclr2019}. We evaluated our method against the five methods reported there: (1) \textit{Semi-supervised Non-negative Matrix Factorization (NMF)~\cite{smaragdis2007supervised}:} The method learns a set of $l=3$ bases from the samples in $\mS_C$ by Sparse NMF~\cite{hoyer2004non, kim2007sparse} as $\mS_C = H_c*W_c$, with mixture components $H_c$ and basis vectors $W_c$, where the two matrices are non-negative, using the fast Non-negative Least Squares solver of~\cite{kim2011fast}. Then, the mixture $\mS_A$ is decomposed with $2l$ bases, where the first $l$ bases are simply $W_c$: $\mS_A = H_{ac}*W_c + H_{ab}*W_b$. The estimated components from domain $B$ are then given by: $\bar \mS_B = h_{ab}*W_b$. (2) \textit{GAN:} A masking function $m$ is learned so that after masking, the training mixtures are indistinguishable from the source samples by a discriminator $d$, similar to our $\mathcal{L}_{\text{GAN}_C}$ loss. (3) \textit{GLO Masking (GLOM):} This method learns an explicit generative GLO~\cite{bojanowski2017optimizing} model to both domain $A$ and domain $C$ and fits the parameters to each given sample $\va$, followed by approximating the solution by a mask between $0$ and $1$ that is multiplied by the mixed signal $\va$. (4) \textit{Neural Egg Separation (NES):} The iterative method of~\cite{yedidiclr2019}, which is initialized by taking the mixture components to be each half of the mixed signal $\va$. (5) \textit{Fine-tuned NES (NES-FT):} Initializing NES with the GLOM solution above. 

The semi-supervised experiments are performed on the MUSDB18~\cite{musdb18} dataset, which consists of $150$ music tracks, $100$ of which in the train set and $50$ in the test set. Each music track is comprised of separate signal streams of the mixture, drums, bass, the accompaniment, and the vocals. In our method, samples are preprocessed as described in Sec.~\ref{sec:prep} and then trained using the method of Sec.~\ref{sec:method}. We compare the performance of our method, using the signal-to-distortion ratio (SDR) in Tab.~\ref{tab:semi}. We can observe that NMF, GAN, GLOM and NES perform much worse then NES-FT and our method. There is also a significant gap between NES-FT and our method ($2.1dB$ vs $3.2dB$) as well. 


\subsection{Comparison to fully supervised methods}
We next compare with fully supervised methods that solely deal with singing voice separation. For this comparison, our evaluation protocol follows closely the one of~\cite{madtwin}, except that our method does not employ the training samples of the singing voices and is unaware of the matching pairs $(\va,\vc)$. Baseline results are shown for \textit{GRA3~\cite{gra3}}, \textit{GRU-RIS-L~\cite{mimilakis2018monaural}} and \textit{MaDTwinNet~\cite{madtwin}}, which are discussed in Sec.~\ref{sec:prev}, and for the following methods: \textit{CHA~\cite{chandna2017monoaural}}, which uses CNN to estimate time-frequency soft masks;   \textit{STO2~\cite{sto2}}, which is based on signal representation that divides the complex spectrogram into a grid of patches of arbitrary sizes; and \textit{JEO2~\cite{jeo2}:}, which is based on robust principal component analysis (RPCA). The results for all of the above approaches are obtained from~\cite{madtwin}.

The development subset of of Demixing Secret Dataset (DSD100)~\cite{dsd100} and the non-bleeding/non-instrumental stems of MedleydB~\cite{medleydb} are used for training. Baseline approaches here are trained in a supervised fashion, while our method is trained in a semi-supervised manner. For evaluation, the evaluation subset of DSD100, which consists of 50 samples, is used. For these methods, the literature reports both the signal-to-distortion ratio (SDR) and the signal to-interference ratio (SIR), and we report both, using the mir\_eval Python library. 

The comparison is shown in Tab.~\ref{tab:sup}. As can be seen, SDR values for our method are better then those of GRA3 and CHA, but worse than STO2, JEO2, GRU-RIS-L and MaDTwinNet. Our SIR value is significantly higher than all baselines, achieving a gap of $7.0$ to the second best method. This is consistent with our observation: The network seems to filter out all the instrumental music very well for most samples. However, for some samples, there is a slight distortion of the voice generated. Samples, in comparison to those published by~\cite{mimilakis2018monaural}, are available at~\url{https://sagiebenaim.github.io/Singing/}.

\subsection{Ablation study}

We perform an ablation analysis to understand the relative contribution of the different losses in our method. This is done by removing various losses from the training objective and retraining. 

As can be seen in Tab.~\ref{tab:ablation}, $\mathcal{L}_{R_2}$ has a smaller significance than other losses. The most significant losses are $\mathcal{L}_{R_4}$, and the GAN losses $\mathcal{L}_{GAN_C}$ and $\mathcal{L}_{GAN_A}$, without even one of these, the two metrics drop considerably. $\mathcal{L}_{R_3}$ is also very significant and without it the SDR is greatly diminished.

\begin{table}[t]
  \caption{Median SDR (dB) for our method and previous semi-supervised approaches evaluated on the MUSDB18~\cite{musdb18} dataset. Baselines are form~\cite{yedidiclr2019}, which did not report SIR.\label{tab:semi}}
  \centering
  \begin{tabular}{cc}
\begin{tabular}{lcc}
\toprule
Approach & SDR & SIR\\
\midrule
NMF & 0.0 &-\\
GAN & 0.3 &-\\
GLOM & 0.6 &-\\
\bottomrule 
\end{tabular} &
\begin{tabular}{lcc}
\toprule
Approach & SDR & SIR \\
\midrule
NES & 0.3 & - \\
NES-FT & 2.1 & - \\
Ours & 3.2  & 14.2\\
\bottomrule 
\end{tabular}
\end{tabular}
  \caption{Median SDR and SIR (dB) values for the proposed method and previous supervised approaches, which solely deal with singing voice separation, evaluated on the evaluation subset of DSD100~\cite{dsd100} dataset.\label{tab:sup}}
  
  \centering
\begin{tabular}{lccc}
\toprule
Approach & Supervision & SDR & SIR \\
\midrule
GRA3~\cite{gra3} & supervised & -1.7 & 1.3\\
CHA~\cite{chandna2017monoaural} & supervised & 1.6 & 5.2\\
STO2~\cite{sto2} & supervised & 3.9 & 6.7\\
JEO2~\cite{jeo2} & supervised & 4.1 & 6.1\\
GRU-RIS-L ~\cite{mimilakis2018monaural} & supervised & 4.2 & 7.9\\
MaDTwinNet~\cite{madtwin} & supervised &  4.6 & 8.2\\
\midrule
Ours & semi-supervised & 3.5 & 15.2 \\
\bottomrule 
\end{tabular}
  \caption{Ablation study: Median SDR and SIR values for the proposed method without (w/o) selected losses evaluated on the evaluation subset of DSD100~\cite{dsd100}.\label{tab:ablation}}
  
  \centering
\begin{tabular}{@{}c@{~~~}c@{}}
\begin{tabular}{l@{~~}c@{~~}c}
\toprule
Losses & SDR &  SIR\\
\midrule
  All losses & 3.5 & 15.2 \\
  w/o $\mathcal{L}_{R_1}$ & -0.9&3.4  \\
  w/o  $\mathcal{L}_{R_2}$  & 2.3 & 9.7 \\
  w/o $\mathcal{L}_{R_3}$ & -4.3 & 13.3\\
 \bottomrule 
\end{tabular} &
\begin{tabular}{l@{~~}c@{~~}c@{}}
\toprule
Losses & SDR &  SIR\\
\midrule
  w/o $\mathcal{L}_{R_4}$ & -6.3&-4.7\\
 w/o $\mathcal{L}_{GAN_A}$  & -6.3 & -4.2 \\
  w/o $\mathcal{L}_{GAN_C}$  & -4.1 & -2.4 \\
  w/o $\mathcal{L}_{GAN_A}$\&$\mathcal{L}_{GAN_C}$ & -17.0 & -3.6 \\
\bottomrule 
\end{tabular} 
\end{tabular}
\end{table}

\section{Conclusions}

We present a new method for semi-supervised singing voice separation that is competitive with some of the state of the art supervised methods and all of the literature semi-supervised ones. The crux of the method is the use of compound losses, applied to synthetic mixes, and the application of two GANs. This setup could be extended to multiple sources due the superposition principle of audio signals that is satisfied by the compound losses and will be inspected as future work. In addition, using time-domain architectures can be explored. The method is applied sequentially to fixed-length audio clips and, as future work, we would like to employ overlapping segments and even incorporate longer-term dependencies.

\subsection*{Acknowlegments}
This project has received funding from the European Research Council (ERC) under the European Unions Horizon 2020 research and innovation programme (grant ERC CoG 725974). 
\bibliographystyle{IEEEbib}
\bibliography{sep_short}

\end{document}